\documentclass[showpacs,preprintnumbers,amsmath,amssymb,aps]{revtex4}

\begin{document}

\title{Anisotropic higher derivative gravity and inflationary universe}
\author{
W. F. Kao\thanks{%
wfgore@cc.nctu.edu.tw} \\
Institute of Physics, Chiao Tung University, Hsinchu, Taiwan}

\begin{abstract}
Stability analysis of the Kantowski-Sachs type universe in pure
higher derivative gravity theory is studied in details. The
non-redundant generalized Friedmann equation of the system is
derived by introducing a reduced one dimensional generalized KS
type action. This method greatly reduces the labor in deriving
field equations of any complicate models. Existence and stability
of inflationary solution in the presence of higher derivative
terms are also studied in details. Implications to the choice of
physical theories are discussed in details in this paper.
\end{abstract}

\maketitle

\section{Introduction}
Inflationary theory is a nice resolution for the the flatness,
monopole, and horizon problems of our present universe described
by the standard big bang cosmology \cite{inflation}. Our universe
is homogeneous and isotropic to a very high degree of precision
\cite{data,cobe}. Such a universe can be described by the well
known Friedmann-Robertson-Walker (FRW) metric \cite{book}.

One expects that gravitational physics could be different from the
standard Einstein models near the Planck scale
\cite{string,scale}. For example, quantum gravity or string
corrections could lead to interesting cosmological
applicastions\cite{string}. Indeed, some investigations have
already addressed the possibility of deriving inflation from
higher order gravitational corrections
\cite{jb1,jb2,Kanti99,dm95}.

In particular, a general analysis of the stability condition for a
variety of pure higher derivative gravity theories could be very
useful in many respects. In fact, it has been shown that a
stability condition should hold for any potential candidate of
inflationary universe in the flat Friedmann-Robertson-Walker (FRW)
space \cite{dm95}. A simple method has been shown useful in such
analysis.

In addition, there is no particular reason why our universe is
initially isotropic to such a high degree of precision. Even
anisotropy can be smoothed out by the proposed inflationary
process, it is also interesting to study the stability of the FRW
space during the post-inflationary epoch. Nonetheless, it is
interesting to study the cases where our universe starts out from
an initially anisotropic universe. As a result, our universe is
expected to evolve from certain anisotropic universe to an stable
isotropic universe. Indeed, it has been shown that there exists
such kind of anisotropic solution for a NS-NS model with a metric
field, a dilaton and an axion field \cite{CHM01}. This
inflationary solution is also shown to be stable against small
field perturbations \cite{ck01}. Note also that stability analysis
has also been studied in various fields of interest
\cite{kim,abel}.

Higher derivative terms should be important for physics near the
Planck scale \cite{kim,dm95}. For example, higher order
corrections from quantum gravity or string theory have been
considered in the studies during the inflationary phase
\cite{green}. Higher derivative terms also arise as the quantum
corrections to the matter fields \cite{green}. Moreover, the
stability analysis of the pure higher derivative gravity models
has been shown in Ref. \cite{dm95}. It is then interesting to
study the implication of this stability analysis in different
models.

Recently, there are growing interests in the study of
Kantowski-Sachs (KS) type anisotropic universes\cite{BPN,LH,NO}.
Hence we will try to study the existence and stability of an
inflationary de Sitter final state in the presence of higher
derivative gravity theory on the Kantowski-Sachs spaces. In
particular, it will be applied to study a large class of pure
gravity models with inflationary KS$/$FRW solutions in this paper.
Any KS type solution that leads itself to an asymptotic FRW metric
at time infinity will be referred to as the KS$/$FRW solution in
this paper for convenience.

It will be shown that the existence of a stable de Sitter
background is closely related to the choices of the coupling
constants. We will try to generalize the work in Ref.
\cite{kp91,kpz99} in order to obtain a general and
model-independent formula for the non-redundant field equations in
the Kantowski-Sachs (KS) type anisotropic space. This equation can
be applied to provide an alternative and simplified method to
obtain the stability conditions in pure gravity theories. In fact,
this general and model-independent formula for the non-redundant
field equations is also very useful in many area of interests.

We will first derive a stability equation which turns out to be
identical to the stability equation for the existence of the
inflationary de Sitter solution discussed in Ref.
\cite{dm95,kpz99}. Note that an inflationary de Sitter solution in
pure gravity models is expected to have one stable mode and one
unstable mode for the system to undergo inflation with the help of
the stable mode. Consequently, the inflationary era will come to
an end once the unstable mode takes over after a brief period of
inflationary expansion. The method developed in Ref.
\cite{dm95,kpz99} was shown to be a helpful way in choosing
physically acceptable model for our universe. Our result
indicates, however, that the unstable mode will also tamper the
stability of the isotropic space. To be more specific, if the
model has an unstable mode for the de Sitter background
perturbation with respect to isotropic perturbation, this unstable
mode will also be unstable with respect to any anisotropic
perturbations.

In particular, we will show in this paper that the roles played by
the higher derivative terms are dramatically different in the
inflationary phase of our physical universe. First of all, third
order term will be shown to determine the expansion rate $H_0$ for
the inflationary de Sitter space. The quadratic terms will be
shown to have nothing to do with the expansion rate of the
background de Sitter space. They will however affect the stability
condition of the de Sitter phase. Their roles played in the
existence and stability condition of the evolution of the de
Sitter space are dramatically different.

\section{Non-redundant field equation and Bianchi identity in KS space}

Given the metric of the following form:
\begin{equation}
ds^2=- dt^2+c^2(t)dr^2 + a^2(t) ( d^2 \theta +f^2(\theta) d
\varphi^2)
\end{equation}
with $f(\theta)= (\theta, \sinh \theta, \sin \theta)$ denoting the
flat, open and close anisotropic space known as Kantowski-Sachs
type anisotropic spaces. To be more specific, Bianchi I (BI),
III(BIII), and Kantowski-Sachs (KS) space corresponds to the flat,
open and closed model respectively. One can instead write the
metric as
\begin{equation} \label{metric}
ds^2=- dt^2+a^2(t)({dr^2\over {1-kr^2}}+r^2d\theta ^2) + a_z^2(t)
dz^2
\end{equation}
with $r$, $\theta$, and $z$ read as the polar coordinates and $z$
coordinate for convenience. One writes it this way in order to
make the comparison with the FRW metric easier. Note that
$k=0,1,-1$ stands for the flat, open and closed universes similar
to the FRW space.

In addition, one can restore the $g_{tt}$ component $b^2(t)$ for
the purpose of deriving the non-redundant field equation
associated with $G_{tt}$ that will be shown shortly. As a result,
one has
\begin{equation}  \label{metricb}
ds^2=-b^2(t) dt^2+a^2(t)({dr^2\over {1-kr^2}}+r^2d\theta ^2) +
a_z^2(t) dz^2
\end{equation}

One can show that all non-vanishing spin connections read
\begin{eqnarray}
\Gamma^t_{tt} &=& H_0 \equiv {\dot{b} \over b}=-{\dot{B} \over 2 B} \\
\Gamma^t_{ii} &=& BH_i g_{ii} \\
\Gamma^i_{ti} &=& H_i  \\
\Gamma^r_{rr} &=& {k r \over 1-kr^2} \\
\Gamma^r_{\theta \theta} &=& -r (1-kr^2) \\
\Gamma^\theta_{r \theta} &=& {1 \over r}.
\end{eqnarray}
Here $B \equiv 1/b^2$ and $H_i \equiv ( \dot{a} /a, \dot{a} /a,
\dot{a_z} /a_z)$ $\equiv (H_1,H_2=H_1, H_z)$ for $r, \theta$, and
$z$ component respectively. One can also define $\Gamma_\mu \equiv
\Gamma_{\mu \nu}^\nu$ and $\Gamma^\mu \equiv \Gamma^{\mu}_{\nu
\alpha} g^{\nu \alpha}$ for convenience. As a result, one can
compute all their non-vanishing components, with $b=1$ being reset
to unity:
\begin{eqnarray}
\Gamma^t &=& 3 H \\
\Gamma^r &=& {1 \over r a^2}  \\
\Gamma_{t} &=& 3H \\
\Gamma_{r} &=& {1 \over r(1-kr^2) } .
\end{eqnarray}

Writing $H_{\mu \nu} \equiv G_{\mu \nu} -T_{\mu \nu}$, one can
show that $D_\mu H^{\mu \nu}=0$ follows from the Bianchi identity
$D_\mu G^{\mu \nu}=0$ and the energy momentum conservation $D_\mu
T^{\mu \nu}=0$ for any energy momentum tensor $T^{\mu \nu}$
coupled to the system. With the metric (\ref{metric}), one can
show that the $r$ component of the equation $D_\mu H^{\mu \nu}=0$
implies that
\begin{equation}
H^r_{\; r}=H^\theta_{\;\theta}.
\end{equation}
The result says that any matter coupled to the system has the
property that $T^r_{\;r}=T^\theta_{\;\theta}$. In addition, the
the equations $D_\mu H^{\mu \theta}=0$ and $D_\mu H^{\mu z}=0$
both vanish identically all by itself irrelevant to the form of
the energy momentum tensor. More interesting information comes
from the $t$ component of this equation. It says:
\begin{equation}
(\partial_t + 3 H) H^t_{\; t} = 2 H_1 H^r_{\; r} +H_zH^z_{\; z}.
\end{equation}
This equation implies that (i) $H^t_{\; t}=0$ implies that
$H^r_{\; r}=H^z_{\; z}=0$ and (ii) $H^r_{\; r}=H^z_{\; z}=0$ only
implies that $(\partial_t + 3 H) H^t_{\; t} =0$. Case (ii) can be
solved to give $ H^t_{\; t} =$ constant $\exp[-a^2a_z]$ which
approaches zero when $a^2a_z \to \infty$. Therefore, for the
anisotropic system one is considering here, the metric contains
two independent variables $a$ and $a_z$ while the Einstein field
equations have three non-vanishing components: $H^t_{\; t} =0$,
$H^r_{\; r}=H^\theta_{\;\theta} =0$ and $H^z_{\; z} =0$. The
Bianchi identity implies that the $tt$ component is not redundant
which needs to be reserved for complete analysis. One can freely
ignore one of the $rr$ or $zz$ components without affecting the
final result of the system. In short, the $H^t_t=0$ equation,
known as the generalized Friedman equation, is a non-redundant
field equation as compared to the $H^r_r=0$ and $H^z_z=0$
equations.

In practice, the field equations of the system may be very
complicate. One of the resolution is to reduce the Lagrangian of
the system from a functional of the metric $g_{\mu \nu}$, ${\cal
L}(g_{\mu \nu})$, to a simpler function of $a(t)$ and $a_z(t)$,
namely $L(t) \equiv a^2 a_z {\cal L}(g_{\mu \nu}(a(t), a_z(t)))$ .
The resulting equation of motion should be able to be
reconstructed from the variation of the reduced Lagrangian $L(t)$
with respect to the variable $a$ and $a_z$. The result is,
however, incomplete because, the variation of $a$ and $a_z$ are
related to the variation of $g_{rr}$ and $g_{zz}$ respectively.
One can never derive the field equation for $g_{tt}$ without
restoring the variable $b(t)$ in advance. This is the reason why
one wishes to introduce the metric (\ref{metricb}) such that the
reduced Lagrangian $L(t) \equiv a^2 a_z {\cal L}(g_{\mu
\nu}(b(t),a(t), a_z(t)))$ contains the non-redundant information
of the $H^t_t=0$ equation. One can reset $b=1$ after the variation
of $b(t)$ has been done. This will reproduce the wanted and
non-redundant Friedman equation.

After some algebra, one can also compute all non-vanishing
components of the curvature tensor:
\begin{eqnarray}
R^{ti}_{\;\;\;tj} &=& [{1\over 2}\dot{B}H_i+B ( \dot{H}_i+H^2_i) ] \delta^i_j \\
R^{ij}_{\;\;\;kl} &=& BH_iH_j \; \epsilon^{ijm}\epsilon_{klm}+{k
\over a^2} \epsilon^{ijz}\epsilon_{klz}.
\end{eqnarray}
Given a Lagrangian $L = \sqrt{g} {\cal L}=L(b(t), a((t), a_z(t))$
one can show that
\begin{eqnarray}
L &=& { a^2 a_z \over \sqrt{B}} {\cal L} (R^{ti}_{\;\;\;tj},
R^{ij}_{\;\;\;kl}) = { a^2 a_z \over \sqrt{B}} {\cal L} (H_i,
\dot{H}_i, a^2)
\end{eqnarray}
The variational equations for can be shown to be:
\begin{eqnarray} \label{key0}
{\cal L} +H_i ( {d \over dt} +3H )L^i &=& H_iL_i + \dot{H}_i L^i \\
{\cal L} + ( {d \over dt} +3H )^2 L^i &=& ( {d \over dt} +3H )L_i
-a^2 { \delta {\cal L} \over  \delta a^2} (1-\delta_{iz})
\end{eqnarray}
Here $L_i \equiv \delta {\cal L} /\delta H_i$,  $L^i \equiv \delta
{\cal L} /\delta {\dot H}_i$, and $3H \equiv \sum_i H_i$. For
simplicity, we will write ${\cal L}$ as $L$ from now on in this
paper. As a result, the field equations can be written in a more
comprehensive form:
\begin{eqnarray} \label{key}
L +H_i ( {d \over dt} +3H )L^i &=& H_iL_i + \dot{H}_i L^i \\
L + ( {d \over dt} +3H )^2 L^i &=& ( {d \over dt} +3H )L_i -a^2 {
\delta {\cal L} \over  \delta a^2} (1-\delta_{iz}) \label{zeq}
\end{eqnarray}


\section{FRW space as a stable final state}
For simplicity, one will start with Einstein-Hilbert (EH) action
and study its evolutionary process from an anisotropic space. It
is known that our final universe is isotropic to a very high
precision. Therefore, any physical model should carry any physical
universe from initial anisotropic space to an isotropic space as a
final destination. Since lowest order Einstein-Hilbert action is
the most well-known popular model, one expects such isotropilized
process should be realized in this model. Any acceptable higher
order terms being considered as corrections around this stable EH
background should not affect its intention evolving toward the FRW
space. Therefore, one will start with a simple EH action with a
cosmological constant term given by:
\begin{equation}
S_{\rm EH}=\int d^4 x \sqrt{g} { \cal L} =\int d^4 x \sqrt{g}
[-R-2\Lambda ] \label{EH} .
\end{equation}
One can show directly that the reduced Lagrangian $L$ is given by
\begin{equation}
L= 4 \dot{H}_1 +2\dot{H}_z+6H_1^2+2H_z^2+ 4H_1H_z +{2k \over
a^2}-2 \Lambda.
\end{equation}
Therefore, one can show that the Friedmann equation (\ref{key})
and $z$-equation (\ref{zeq}) take the following form:
\begin{eqnarray} \label{keyEH}
&& H_1^2+ 2 H_1H_z+ {k \over a^2}= \Lambda, \\
&& 2 \dot{H}_1+ 3 H_1^2 + {k \over a^2}= \Lambda. \label{zeqEH}
\end{eqnarray}
For convenience, one can also use equations
(\ref{keyEH}-\ref{zeqEH}) to derive the following equation:
\begin{eqnarray}
&& \dot{H}_1+ H_1^2 = H_1H_z. \label{eq2}
\end{eqnarray}
Eq. (\ref{eq2}) can be shown to give
\begin{equation}
a_z= k_0 \dot{a}
\end{equation}
for some integration constant $k_0$. Note also that Eq.
(\ref{zeqEH}) can also be integrated as
\begin{equation}
H_1^2+{ k \over a^2}={\Lambda \over 3} + { k_1 \over a^3}
\end{equation}
for some integration constant $k_1$. For the case $k=0$, one can
show that $\ddot{A}=9H_0^2A/4$ if we write $A = a^{3/2}$.  Hence
one can integrate this equation to obtain
\begin{equation}
a(t)= a(0) \left [ { \exp [3H_0t/2] +k_2  \exp [-3H_0t/2] \over
1+k_2} \right ]^{2/3}
\end{equation}
for come constant $k_2$. Here $H_0^2 =\Lambda/3$ denotes the
expansion factor. For the case where $k \ne 0$, a special solution
with $k_1=0$ can be found to be
\begin{eqnarray}
&& a=a_1 \left [  \exp [H_0t] + {k \over 4H_0^2a_1^2} \exp [-H_0t]
\right ]
\end{eqnarray}
In fact, one can show that the evolutionary properties of these
solutions can be obtained without knowing the exact solutions.
Indeed, one will show in a moment that the inflationary solution
will try to evolve to an isotropic FRW space as $t \to \infty$. In
addition, one can also show that these solution will remain
isotropic from an stability analysis. One will study the evolution
of physical universe from anisotropic initial state to isotropic
final state. Therefore, one will write above equations in terms of
the following variables:
\begin{eqnarray} \label{v1}
&& 3H = 2H_1+ H_z ={ \dot{V} \over V}, \\
&& \Delta \equiv (H_1-H_z)  \label{v2}
\end{eqnarray}
with $\Sigma$, $V=a^2a_z$, and $\Delta$ the total expansion rate,
3-volume, and the deviation function respectively. One expects
$\Delta \to 0$ as the physical universe evolves toward an
isotropic final state. Indeed, one can show that the field
equations can be written as
\begin{eqnarray}
&& \dot{\Delta}+ 3H \Delta = -{k \over a^2}. \label{v33}
\end{eqnarray}
Eq. (\ref{v33}) can be rewritten as
\begin{equation}
{d \over dt} (V\Delta) =-k a_z.
\end{equation}
Hence one has
\begin{equation}
 \Delta = \Delta(0){V(0) \over V} - {k a^2(0) \over a^2 }{\int_0^t
 a_z(t')dt' \over a_z}
\end{equation}
If $a_z$ eventually expands proportional to $\exp [\Lambda_0 t]$,
one can show that ${\int_0^t
 a_z(t')dt'/ a_z} \to 1 /\Lambda_0$, a small constant, as $t \to
 \infty$. Since the scaling factor $a^2(0)/ a^2 \to 0$ for an
 expanding $a$ solution, this equation implies that $\Delta \to 0$
 as $t \to \infty$. Hence one shows that the field equations of
 the EH action will definitely take the anisotropic universe,
 either one of the KS type spaces, to the final FRW universe as $t
 \to \infty$.

One can also show that the final isotropic FRW universe is stable
against any small perturbations $H_1=H_0 + \delta H_1$ and
$H_z=H_0+\delta H_z$. For convenience, one can also use equations
(\ref{keyEH}-\ref{zeqEH}) to derive the following equation:
\begin{eqnarray} \label{eq1}
&& \dot{H}_z +H_z^2+ 2 H_1H_z = \Lambda.
\end{eqnarray}
Applying this perturbations to Eq.s (\ref{eq2}) and (\ref{eq1}),
one has
\begin{eqnarray}\label{st2}
&&  \delta \dot{H}_1 +H_0 \delta H_1 -H_0 \delta H_z=0, \\
\label{st1} && \delta \dot{H}_z +4H_0 \delta H_z + 2H_0 \delta
H_1=0 .
\end{eqnarray}
Adding equations (\ref{st2}-\ref{st1}) one can derive
\begin{eqnarray}
&&  (\delta \dot{H}_1 +3H_0 \delta H_1 ) + (\delta \dot{H}_z +3H_0
\delta H_z ) =0. \label{st23}
\end{eqnarray}
Hence one has $\delta H_1+\delta H_z =$ constant$/a^3 \to 0$ as $a
\to \infty$. Hence any physical perturbation against the the FRW
background would imply $\delta H_1 \to -\delta H_z \equiv \delta
H$. Hence, Eq. (\ref{st2}) implies
\begin{equation}
\delta \dot{H} +2H_0 \delta H =0.
\end{equation}
This equation can be integrated to obtain the result
\begin{equation}
\delta H_1= {{ \rm constant} \over a^2} \to 0
\end{equation}
as $t \to \infty$. Hence one can again show that the final FRW
space is stable against this anisotropic perturbation. Note that
both $\delta H_i \to 1/a^2$ asymptotically, while their signs in
the order of $O(a^{-2})$ are opposite to each other such that
their sum $\delta H_1+ \delta H_z \to 1/a^3$.

\section{Stability of higher derivative inflationary solution}
One can then apply the perturbation, $H_i=H_{i0}+ \delta H_i$, to
the field equation with $H_{i0}$ the background solution to the
system. This perturbation will enable one to understand whether
the background solution is stable or not. In particular, one would
like to learn whether a KS $\to$ FRW (KS/FRW) type evolutionary
solution is stable or not.

Note that our universe could start out anisotropic even evidences
indicate that our universe is isotropic to a very high degree of
precision in the post inflationary era. Therefore, one expects
that any physical model should admit a stable KS$/$FRW solution.
In particular, one will be interested in a de Sitter (dS)
background solution with $H_{i0}=H_0$ for some constant Hubble
expansion parameter. One will denote such solution as KS $\to$ de
Sitter (KS/dS) type inflationary solution.

One can show that any FRW inflationary solutions with a stable
mode and an unstable mode is a negative result to our search for a
stable inflationary model. In particular, any FRW inflationary
solutions with a stable mode and an unstable mode will provide a
natural way for the inflationary universe to exit the inflationary
phase. Such models will, however, also be unstable against the
anisotropic perturbations. Therefore, such solution will be
harmful for the system to settle from anisotropic space to FRW
space once the graceful exit process is done. One will show in
this section that the higher derivative gravity theory one
considers here could also accommodate two stable modes with
appropriately chosen coupling constants. In such case, the
inflationary de Sitter solution $H_0$ will also be stable against
anisotropic perturbations.

First of all, one can show that the first order perturbation
equation from the non-redundant field equation (\ref{key}), with
$H_i \to H_0+\delta H_i$, gives
\begin{eqnarray}\label{stable0}
&& <H_i L^{ij} \delta  \ddot{H}_j> +3H <H_i L^{ij} \delta
\dot{H}_j> + \delta <H_i\dot{L}^i>+3H <(H_i L^i_j + L^j) \delta
H_j> \nonumber \\
&& +<H_iL^i> \delta (3H) =<H_iL_{ij} \delta
H_j>+<\dot{H}_i \delta L^i> \label{stable1}
\end{eqnarray}
with all functions of $H_i$ evaluated at some FRW background with
$H_i=H_0$. The notation $< A_iB_i> \equiv \sum_{i=1,z} A_iB_i$
denotes the summation over $i=1$ and $z$ for repeated indices.
Note that we have absorbed the information of $i=2$ into $i=1$
since they contributes equally to the field equations in the KS
type spaces. In addition, $L^{i}_{j} \equiv \delta^2 L / \delta
\dot{H}_{i} \delta H_j$ and similarly for $L_{ij}$ and $L^{ij}$
with upper index $^i$ and lower index $_j$ denoting variation with
respect to $\dot{H}_i$ and $H_j$ respectively for convenience. In
addition, perturbing Eq. (\ref{zeq}) can also be shown to
reproduce the Eq. (\ref{stable0}) in the FRW
limit\cite{inflation}.

Once we adopt the de Sitter solution with $H_0=$ constant, one has
\begin{eqnarray}
<H_i L^{ij} \delta  \ddot{H}_j> +3H <H_i L^{ij} \delta \dot{H}_j>
+3H <(H_i L^i_j + L^j) \delta H_j> +<H_iL^i> \delta (3H)
=<H_iL_{ij} \delta H_j> \label{stable1}
\end{eqnarray}
where all field variables are understood to be evaluated at the
background de Sitter space where $H_i = H_0 = $ constant for all
directions.

If the inflationary de Sitter solution has one one stable mode and
one unstable mode for the system, the unstable mode is expected to
collapse the de Sitter phase. Then the inflationary era will come
to an end once the unstable mode takes over. It was shown earlier
to be a helpful way to select physically acceptable model for our
universe. Our result shown here indicates, however, that the
unstable mode will also tamper the stability of the isotropic
space. Indeed, if the model has an unstable mode for the de Sitter
perturbation, this unstable mode will also be unstable against the
anisotropic perturbation.

For example, one can show that the model \cite{dm95}
\begin{equation}
{\cal L} = -R + \alpha R^2 + \beta R^\mu_\nu R^\nu_\mu + \gamma
R^{\mu \nu}_{\;\;\; \beta \gamma} \, R^{\beta \gamma}_{\;\;\;
\sigma \rho} \, R^{\sigma \rho}_{\;\;\; \mu \nu}
\end{equation}
admits an inflationary solution if $\gamma >0$. Note that the
$\gamma$ term is the minimal consistent effective low-energy
two-loop renormalizable Lagrangian for pure gravity theory
\cite{2loop}. In addition, the quadratic terms can be shown to be
derivable from the matter effect of quantum fields. For
simplicity, one can write
\begin{eqnarray}
&& A=\dot{H}_1+H_1^2, \\
&& B= H_1^2 + {k \over a^2}, \\
&& C=H_1H_z, \\
&& D=\dot{H}_z+H_z^2
\end{eqnarray}
since all curvature tensor components and all field equations will
be functions of above combinations. This notation will be shown to
very convenient in tracking the field equations for any complicate
models such as the higher derivative models we are working on in
this paper. Indeed, one can show that the Lagrangian reads
\begin{eqnarray}&&
L= 4A+2B+4C+2D+ 4 \alpha
\left[4A^2+B^2+4C^2+D^2+4AB+8AC+4AD+4BC+2BD+4CD \right] \nonumber
\\ && + 2 \beta \left[3A^2+B^2+3C^2+D^2+2AB+2AC+2AD+2BC+2CD \right]
+8\gamma \left[2A^3+B^3+2C^3+D^3 \right].
\end{eqnarray}
This Lagrangian will reduce to the de Sitter models when we set
$H_i \to H$ in the isotropic limit. Hence one can show that the
generalized Friedmann equation (\ref{key}) gives
\begin{equation} \label{gamma1}
H_0^4 = 1/4 \gamma
\end{equation}
in the isotropic limit for an inflationary solution with
$\dot{a}^2 \gg 1$.

Note that the quadratic terms does not contribute to the Friedmann
equation in the de Sitter limit with a constant inflationary
phase. This can be shown to be a general property of the quadratic
models. One will show later that these terms will, however, affect
the stability conditions for the inflationary phase. Indeed,
quadratic terms will be shown to affect the duration of the
inflationary de Sitter phase which will be shown in a moment.

Note that one can also show that
\begin{eqnarray}
<H_i L^{i1}> &=& 2 <H_i L^{iz}  > ,\\
<H_i L^i_1> &=& 2 <H_i L^i_z>, \\
L^1 &=& 2 L^z ,\\
<H_iL_{i1}> &=& 2 <H_iL_{iz}>,
\end{eqnarray}
in the inflationary de Sitter background with $H_0=$ constant.
Therefore, the stability equations (\ref{stable1}) can be greatly
simplified. For convenience, one will define the operator ${\cal
D}$ as
\begin{eqnarray}
{\cal D} \delta H \equiv  <H_i L^{i1}> \delta  \ddot{H} +3H <H_i
L^{i1}> \delta \dot{H} +3H <H_i L^i_1 + L^1>  \delta H+2 <H_iL^i>
\delta H - <H_iL_{i1}> \delta H.
\end{eqnarray}
As a result, one can show that the stability equation
(\ref{stable1}) reads
\begin{equation}
{\cal D} (\delta H_1+ \delta H_z)=0
\end{equation}
with $i=1$ and $i=z$ component summed over. Since, one can perturb
the filed $H_i$ from any directions, stability conditions must
also hold for perturbation from any directions. Hence, one expects
the stability conditions must hold and be identical from the
perturbations in each directions. Indeed, above result shows that
both $\delta H_1$ and $\delta H_z$ follows the same condition.
Hence one needs to solve the following stability condition ${\cal
D} \delta H_j=0$ for both $i=1$ and $i=z$. Hence one has
\begin{eqnarray}
<H_i L^{i1}> \delta  \ddot{H}_j +3H <H_i L^{i1}> \delta \dot{H}_j
+3H <H_i L^i_1 + L^1>  \delta H_j+2 <H_iL^i> \delta H_j =
<H_iL_{i1}>  \delta H_j\label{stable11}
\end{eqnarray}
to see if the system is stable or not against any small
perturbation with respect to the de Sitter background.

In addition, the stability equation (\ref{stable11}) for $\delta
H_i$ can be shown to be
\begin{equation}
\left [ 6 \alpha + 2 \beta + 12 \gamma H_0^2 \right ] \left [
\delta \ddot H_i  + 3H_0 \delta \dot H_i \right ]  +(1- 12 \gamma
H_0^4 ) \delta H_i =0
\end{equation}
for such KS/dS solutions. Hence one has
\begin{equation}
\delta H_i = c_i \exp [{-3H_0t \over 2} (1+ \delta_1) ] + d_i \exp
[{-3H_0t \over 2} (1- \delta_1) ]
\end{equation}
with
\begin{equation}
\delta_1 = \sqrt{1+ 8/[27+9 (6 \alpha+2 \beta)H_0^2 ] }
\end{equation}
and some arbitrary constants $c_i, d_i$ to be determined by the
initial perturbations. It is easy to see that any small
perturbation $\delta H_i$ will be stable against the the de Sitter
back ground if both modes characterized by the exponents
\begin{equation}
\Delta_{\pm} \equiv - [3H_0t / 2] [1 \pm \delta_1]
\end{equation}
are all negative. This will happen if $\delta_1 < 1$. In such
case, the inflationary de Sitter space will remain a stable
background as the universe evolves.

On the other hand, one would have a stable mode and an unstable
mode if $\delta_1 >1$. This indicates that this model admits one
stable mode and one unstable mode following the stability equation
(\ref{stable11}) for the inflationary de Sitter solution. It is
shown to be a positive sign for an inflationary model that is
capable of resolving the graceful exit problem in a natural
manner.

Indeed, one expects any unstable mode for a model to be of the
form $\delta H_i \sim \exp [lH_0t]$, to the lowest order in
$H_0t$, in a de Sitter background with $l$ some constant
characterizing the stability property of the model. In such
models, the inflationary phase will only remain stable for a
period of the order $\Delta t \sim 1/lH_0$. The inflationary phase
will start to collapse after this period of time. This means that
the de Sitter background fails to be a good approximation when $t
\gg \Delta t$.

As a result, the anisotropy will also grow according to $\delta
H_i \to \delta H_i^0 \exp [lH_0 \Delta t]$ with $\delta H_i^0$
denoting the initial perturbation. Hence this model will have
problem remaining isotropic for a long period of time. Therefore,
pure gravity model of this sort will not solve the graceful exit
problem. One will need, for example, the help of certain scalar
field to end the inflation in a consistent way. The unstable mode
gives us, however, a hope that small anisotropy observed today can
be generated by the initial inflationary instability for models
with appropriate factor $l$.

The result shown in this paper shows that the roles played by the
higher derivative terms are dramatically different in the
inflationary phase of our physical universe. First of all, third
order term characterized by the coupling constant $\gamma$  will
determine the expansion rate $H_0$, given by Eq. (\ref{gamma1}),
for the inflationary de Sitter space. The quadratic terms
characterized by the coefficients $\alpha$ and $\beta$ will not
affect the expansion rate of the background de Sitter space. They
will however affect the stability condition of the de Sitter phase
depending on the sign of the characteristic function
$\Delta_{\pm}$. Both the third order term and quadratic terms are
closely related to the quantum corrections of the quantum
fields\cite{green,2loop}. Their roles played in the existence and
stability condition of the evolution of the de Sitter space are
dramatically different. They are however equivalently important in
the higher derivative models.

\section{conclusion}
We have tried to generalize the work in Ref. \cite{kp91,kpz99} in
order to obtain a general and model-independent formula for the
non-redundant field equations in the Kantowski-Sachs (KS) type
anisotropic space. This equation can be applied to provide an
alternative and simplified method to obtain the stability
conditions in pure gravity theories. In fact, this general and
model-independent formula for the non-redundant field equations is
also very useful in many area of interests.

It is also shown that the existence of a stable de Sitter
background is closely related to the choices of the coupling
constants. We first derive a stability equation which turns out to
be identical to the stability equation for the existence of the
inflationary de Sitter solution discussed in Ref.
\cite{dm95,kpz99}.

If the inflationary de Sitter solution in the pure gravity theory
has one one stable mode and one unstable mode for the system, the
unstable mode is expected to collapse the de Sitter phase.
Consequently, the inflationary era will come to an end once the
unstable mode takes over after a brief period of inflationary
expansion. The method developed in Ref. \cite{dm95,kpz99} is
useful way in choosing physically acceptable model for our
universe. Our result indicates, however, that the unstable mode
will also tamper the stability of the isotropic space.

To be more specific, if the model has an unstable mode for the de
Sitter background perturbation with respect to isotropic
perturbation, this unstable mode will also be unstable with
respect to any anisotropic perturbations. In particular, we have
shown in this paper that the roles played by the higher derivative
terms are dramatically different in the inflationary phase of our
physical universe. First of all, third order term is shown to
determine the expansion rate $H_0$ for the inflationary de Sitter
space. The quadratic terms will be shown to have nothing to do
with the expansion rate of the background de Sitter space. They
will however affect the stability condition of the de Sitter
phase. Their roles played in the existence and stability condition
of the evolution of the de Sitter space are dramatically
different.

In short, the result of this paper shows that graceful exit and
stability of any de Sitter model can not work along in a naive
way. The physics behind the inflationary de Sitter models appears
to be much more complicate than one may expect. In another words,
the phase transition during and after the inflationary phase
deserves more attention and requires extraordinary care in order
to resolve the problem lying ahead.

In addition, the result shown in this paper shows that the roles
played by the higher derivative terms are dramatically different
in the inflationary phase of our physical universe. First of all,
third order term characterized by the coupling constant $\gamma$
will determine the expansion rate $H_0$, given by Eq.
(\ref{gamma1}), for the inflationary de Sitter space. The
quadratic terms characterized by the coefficients $\alpha$ and
$\beta$ will not affect the expansion rate of the background de
Sitter space. They will however affect the stability condition of
the de Sitter phase depending on the sign of the characteristic
function $\Delta_{\pm}$. Both the third order term and quadratic
terms are closely related to the quantum corrections of the
quantum fields\cite{green,2loop}. Their roles played in the
existence and stability condition of the evolution of the de
Sitter space are dramatically different. They are however
equivalently important in the higher derivative models.

\section*{Acknowledgments}

This work is supported in part by the National Science Council of
Taiwan.

\end{document}